\documentclass[prc,twocolumn,tightenlines,nofootinbib]{revtex4-2}

\usepackage{graphicx}
\usepackage{dcolumn}
\usepackage{bm}
\usepackage{amssymb}
\usepackage{lipsum}
\usepackage{mathtools}
\usepackage{epstopdf}
\usepackage{amsmath}
\usepackage{braket}
\usepackage{float}
\newcommand{\mytitle}{\boldmath{Octupole deformation properties in the actinides region using Fayans functionals}\unboldmath}

\begin{document}

\preprint{APS/123-QED}

\title{\protect\mytitle}

\author{Gauthier Danneaux}
\author{Markus Kortelainen}%
\affiliation{%
Department of Physics, University of Jyväskylä, PB 35 (YFL) FIN-40014 Jyväskylä, Finland
}%

\date{\today}

\begin{abstract}
In this first-of-its-kind survey conducted on heavy and deformed nuclei in the actinide region of the nuclear chart, we have charted nuclear ground state properties predicted by Fayans energy density functionals (EDFs), focusing in particularly on octupole deformability. Compared to earlier studies with Skyrme-based EDFs, we found similar region of octupole deformed nuclei. Moreover, Fayans EDFs were found to provide accurate predictions for various ground state properties, when compared to experimental data. Comparison to Skyrme-based EDF shows similar trends in various nuclear properties.
\end{abstract}

\maketitle


\section{Introduction}

In light of the recent theoretical and experimental studies and surveys, it appears that the systematics of many observables of atomic nuclei are unequivocally dependent on local fluctuations~\cite{Myeong2023}, shapes and deformations~\cite{Geldhof2022,Sassarini2022}, or configuration mixing~\cite{Kock1971}. Some these noticeable properties, such as the binding energy and the root-mean-square (rms) charge radius, are crucial elements in order to properly understand and efficiently probe and develop uses for said nuclei. As such, understanding the mechanisms behind these phenomena is one of the major inquiries within theoretical nuclear physics.

Intrinsic multipole deformations, a broad and crucial element in the nuclear Density Functional Theory (DFT), as well as many other properties such as the nuclear radii or the binding energies, can be theoretically investigated with DFT calculations~\cite{Ring1980}, which are based upon a specified Energy Density Functional (EDF)~\cite{Schunck2019}. Comparison of theoretical predictions to experimental results can be used to assess the EDF model's precision and accuracy~\cite{Groote2024}, eventually refining its parameters.

Due to their robust results and relative low computational cost, the most commonly used modern nuclear EDFs are the Skyrme functionals. The various parameters contained within such EDFs can be adjusted to specified experimental inputs, with the goal in mind to attain satisfactory predictive power regarding various observables. New experimental data as well as developments in the Skyrme-based EDFs gave rise to state-of-the-art functionals. Among these EDFs are the widely used UNEDF EDFs, including UNEDF0~\cite{Kortelainen2010}, which have been adjusted by using a large set of nuclei experimental data, spanning the whole nuclear chart and including deformed open-shell nuclei. As results, these Skyrme-based EDFs have been shown to present a good predictive power across various nuclear properties, from ground state energy and multipole deformation factors to nucleon separation energies and nuclear radii across the whole nuclear chart, especially regarding its heavier half~\cite{Bonnard2023}.

Indeed, since the quadrupole spectroscopic moment can be compared with an overall significant agreement with experimental data even for heavy and odd-$A$ nuclei, both in ground and fixed isomeric states~\cite{Bonnard2023}, those UNEDF EDFs have been shown to predict accurately results regarding such moment. However, despite these groundbreaking results, the reflection-asymmetric octupole deformation remains less discussed and studied, both theoretically and experimentally. Such shapes are typically found in the heaviest part of the nuclear chart, for example, within the heavier actinides and trans-actinides. The octupole deformation in these nuclei has an important connection to the nuclear Schiff moment~\cite{Engel2003,Dobaczewski2018} among other properties~\cite{Ebata2017}.

Far from being anecdotal when compared to the quadrupole deformation, static octupole deformation, also called reflection asymmetric shape, actually manifests a profound signature on the nuclear observables~\cite{Flambaum2022,Ramachandran2023} and thus also on the systematics of the nuclear ground state properties and excitation spectra~\cite{Butler2020}. As a result, collective octupole deformation can be extracted from, for example, $E3$ transitions, and its effects can be seen on various nuclear properties such as low-lying negative parity odd state energies in even-even nuclei or alternating-parity rotational bands~\cite{Butler1996}. The presence of reflection-asymmetric, axially symmetric nuclei can be partly attributed to parity doublets within $\Delta l =\Delta j =3$ nucleon single-particle shells, leading to small yet non-negligible reflection-asymmetric deformation shapes, mainly due to the differences between even and odd-components of the multipolar nuclear binding energy~\cite{Chen2021}. Additionally, octupole deformation and reflection asymmetric shapes appear as strongly relevant during the fission process, as demonstrated within microscopic descriptions of such phenomenon including multi-variable potential energy-surface~\cite{Schunck2016}. Therefore, its understanding is crucial for various high-end phenomena such as the role of r-process within translead nucleosynthesis, as the abundance of heavy fission material has been shown to display strong effects on the free neutron density and thus push the freeze-out beyond lead in neutron star mergers~\cite{Giuliani2020}. Moreover, octupole deformation is directly connected on fission fragments mass distribution as asymmetric fission seemingly plays a crucial role in the mass distribution. Such deviation is consistent with both the observation of asymmetric components as well as previous theoretical predictions~\cite{Tripathi2015}. Furthermore, all of these phenomena and properties can also lead to a better understanding of the dynamics and effects of fission, including improved power plants~\cite{Butler2020}.

On the other hand, previous theoretical results have motivated measurement of the octupole deformation of Barium isotopes through their rotational bands~\cite{Brewer2023}. This survey showed that such deformation tends to stabilize the vibrational degrees of freedom, as the evidence of the rotational effect of octupole deformation on even-$Z$ nuclei is put in perspective. Moreover, the experimental deformation energies were found to be in a relatively good agreement with previous theoretical predictions with UNEDF EDFs when octupole deformation is involved in the calculations~\cite{Cao2020}. This showed the importance of octupole deformation in physical nuclei as well as in theoretical calculations. Additionally, in-source resonance ionization spectroscopy at the ISAC facility in TRIUMF also highlighted the presence of octupole deformation, which provided the best agreements with experimental data when included compared to other predictions that did not include this degree of freedom, especially regarding rms radius, where octupole deformation appears to play a crucial role~\cite{Verstraelen2019}. These results further indicate the crucial need for improvements in the theoretical and experimental descriptions of neutron-rich octuple-deformed actinide nuclei. As a striking example, the BSkG EDFs~\cite{Scamps2021} have been adjusted at triaxially-deformed HFB level, including later also octupole degree of freedom~\cite{Grams2023}. Moreover, recent theoretical calculations showed that current Skyrme-based EDFs do not reproduce with satisfactory accuracy some more subtle properties, namely the minute variations found alongside isotopic chains, when compared to experimental results~\cite{Kortelainen2022,Rong2025}. Additionally, a survey showed that a microscopic quadrupole-octupole collective Hamiltonian model can describe the empirical low-energy quadrupole and octupole collective states trend with great accuracy, which were coherent with relativistic and non-relativistic EDF microscopic calculations~\cite{Xia2017}. Finally, recent experiments showed the possibility of direct highlighting of octupole deformation effects in barium isotopes~\cite{Bucher2016} and the indirect measurement of octupole deformation in Radon isotopes using accelerated radioactive beams~\cite{Gaffney2013}, hinting at the feasibility of further studies of octupole-deformed nuclei across the nuclear chart.

As a whole, due to its overall relevance, the impact of octupole deformation on the nuclear systematics and structures, and current challenges with theoretical models' accuracy, the nuclear static octupole moment poses an important test for theoretical nuclear structure models. Overall, an experimentally-accurate reproduction of the local variations and staggering of the nuclear radii and binding energy that can be attributed to a better integration of the presence of octupole deformation, a requirement posing a challenge that the next generation of EDF models is expected to overcome.

Recent developments in the DFT domain have led to new approaches of the nuclear pairing interaction. For a long time, the effective particle–particle interaction usually consisted of a rather rudimentary density-dependent contact interaction form. This pairing interaction was extended in~\cite{Fayans2000} to include a gradient term. Moreover, higher-order terms have also been considered, for example, with EDFs intended for beyond mean-field calculations~\cite{Sadoudi2013}. Following these developments, recent Fayans EDFs~\cite{Reinhard2017} would then include gradient term on the particle-particle channel, thus vastly improving their ability to describe some nuclear properties such as the nuclear charge radius. In essence, this gradient term mimics effective finite range interaction effects.

As shown in previous research, Fayans functionals are able to accurately reproduce many nuclear properties and observables, including alpha-decay $Q$-values without any additional parameter readjustment~\cite{Tolokonnikov2017}. Moreover, recent calculations with Fayans EDFs have been shown to be able to reproduce nuclear charge radius local variations with remarkable accuracy across the whole nuclear chart~\cite{Reinhard2017,Wang2024}, whereas Skyrme-based EDFs had difficulties to reproduce such behavior. As such, new measurements on nuclear charge radii offer one of the significant testing ground for new EDF models. Thanks to recent breakthroughs in experimental techniques, for example, with laser spectroscopy,~\cite{Lynch2018,Verstraelen2019,Koszorus2021,Reponen2021,Wang2024}, new data on the local variation of the charge radius has become available for many isotopic chains across the nuclear chart. This has allowed systematic tests of various theoretical models~\cite{Miller2019,Ohayon2022,Rong2024,Kortelainen2022}.

Based on these recent theoretical developments, indicating the importance of octupole deformation as well as the crucial role of the pairing interaction within EDFs for the prediction of the nuclear charge radii, we have conducted a first-of-its-kind  survey of octupole deformation properties of the Fayans EDF models. Thus, we studied namely the ground state energy, the quadrupole and octupole deformations, and the nuclear charge radius, and compared our results to the current generation of Skyrme-based model, namely the UNEDF0, in order to assess the possible improvements for Fayans EDFs. As such, the goal of this work is to examine the various aspects of the Fayans EDFs connected to octupole deformation and of its consequences on some major nuclear properties in order to assess and compare them to earlier experimental and Skyrme-based theoretical results, and to investigate the overall predictive power of Fayans EDFs, from charge radius to quadrupole deformation and separation energies in the actinide and trans-actinide region.

\section{Theoretical framework}
In the HFB framework, the density matrix and pairing tensors can be used as a building blocks of the theory. They are defined~\cite{Ring1980} as
\begin{equation}
    \rho_{ij}=\braket{\Psi\vert c^\dagger_j c_i\vert\Psi};\  \kappa_{ij}=\braket{\Psi\vert c_j c_i\vert\Psi}\, ,
\end{equation} 
which will be relevant in the subsequent calculations where we use two-body density-dependent Hamiltonian
\begin{equation} \label{eq:1}
    \hat{H}=\Sigma_{ij} t_{ij}c^\dagger_i c_j+\frac{1}{4}\Sigma_{ijkl}\Bar{v}_{ijkl}c^\dagger_ic^\dagger_j c_l c_k \,.
\end{equation}
By utilizing the variational principle, the HFB equations then read as~\cite{Ring1980, Schunck2019}
\begin{equation}\label{eq:hfb_eq}
\begin{pmatrix}
\hat{h}-\lambda&\hat{\Delta}\\
-\hat{\Delta}^*&-\hat{h}^*+\lambda
\end{pmatrix}
\begin{pmatrix}
U_k\\
V_k
\end{pmatrix}=E_k\begin{pmatrix}
U_k\\
V_k
\end{pmatrix},
\end{equation}
where $U,V$ are the unitary matrices of the general Bogoliubov transformation, $h_{ij}=t_{ij}+\sum_{kl}\Bar{v}_{iljk}\rho_{kl}$ is the self-consistent mean-field on particle-hole channel, $\Delta_{ij}=\frac{1}{2}\Sigma_{kl}\Bar{v}_{ijkl}\kappa_{kl}$ is the self-consistent pairing field, and $\lambda$ is the chemical potential required to ensure the correct average particle number.

In fact, when computing nuclear radii, it is important to use the HFB formalism rather than HF+BCS approach. While the latter is more straightforward, the unification of HF and BCS under a single variational theory using a set of generalized states and an approach beyond a direct mean-field approximation allows the HFB formalism to solve the neutron gas problem, present in the HF+BCS treatment~\cite{Dobaczewski1984}. In the case of a straightforward BCS theory, quasiparticle states in continuum are not properly localized, which leads to unrealistic prediction of neutron density outside the nucleus, impacting thus also on the nuclear radius. Thus, the HFB formalism allows for more reliable solutions for neutron-rich nuclei far away from the valley of stability, where deformation, long-ranged pairing effects, and charge radii are more realistically handled.

Nevertheless, as it can be seen in Eq. (\ref{eq:hfb_eq}), this equation is non-linear, as the mean-field $h$ and pairing field $\Delta$ both dependent on the HFB matrices $U$ and $V$, which are at the same time solutions of this equation. This recursion implies that this equation cannot be solved directly and therefore requires an iterative process. This process can be initiated through by an initial guess on the fields, and then in subsequent iterations computing the energy and fields through the use of an EDF, which will then drive the whole process and be the key element regarding the resulting nuclear properties. From the HFB solution, various observables can then be computed accordingly, such as the ground state energy, multipole deformation factors, and so on.

Within nuclear DFT framework, the total binding energy of the nucleus is computed as
\begin{equation}
    E=\int \mathcal{E}(\pmb{r})d\pmb{r}^3\, ,
\end{equation}
where $\mathcal{E}$ is the EDF, a real, time-even and scalar functional encompassing all the considered local currents, densities, and their derivatives.

Both the Skyrme force~\cite{Skyrme1958} --- a straightforward zero-range interaction --- and the Gogny force~\cite{Decharge1980} --- which consists of a finite-range part and a zero-range density-dependent term together with a zero-range spin-orbit interaction --- constitute the majority of present-day non-relativistic nuclear EDFs. The Skyrme EDF, which is usually separated into a time-even and a time-odd parts, relies mainly on local nucleon, kinetic energy, and spin-current densities on its time-even part. In fact, it is due to the simpler form of the Skyrme EDF that its parameter adjustments have been pushed the farthest so that a large variety of applications have been developed for them.

On the other hand, the Fayans EDFs are typically based around the following volume, surface, spin-orbit, Coulomb exchange, and pairing energy densities~\cite{Fayans1994, Fayans1998, Fayans2000, Reinhard2017}:
\begin{eqnarray}
\mathcal{E}^{\rm v}_{\text{Fy}}&=&\frac{1}{3}\epsilon_F\rho_{\text{sat}}\left[a_v^+\frac{1-h_{1+}^vx^\sigma_0}{1+h^v_{2+}x^\sigma_0}x^2_0\right. \nonumber \\
&+&\left.a_v^-\frac{1-h_{1-}^vx^\sigma_0}{1+h^v_{2-}x^\sigma_0}x^2_1\right] \,,\\
\mathcal{E}^{\rm s}_{\text{Fy}}&=&\frac{1}{3}\epsilon_F\rho_{\text{sat}}\frac{a^s_+r^2_s(\nabla x_0)^2}{1+h^s_+ x^\sigma_0+h^s_\nabla r^2_s (\nabla x_0)^2} \,,\\
\mathcal{E}^{\rm ls}_{\text{Fy}}&=&\frac{4\epsilon_F r^2_s}{3\rho_{\text{sat}}}\left(\kappa \rho_0 \nabla\cdot J_0+\kappa'\rho_1\nabla\cdot J_1+g J^2_0\right. \nonumber\\
&+&\left.g' J^2_1\right) \,,\\
\mathcal{E}_{\rm C,ex}&=&-\frac{3}{4}e^2(\frac{3}{\pi})^{1/3}\rho_p^{4/3}(1-h_Cx^\sigma_0)\,,\\
\mathcal{E}^{\text{pair}}_{\text{Fy},q}&=&\frac{2\epsilon_F}{3\rho_{\text{sat}}}{\tilde{\rho}_q}^2[f^\xi_{ex}+h^\xi_+x^\gamma_{\text{pair}}+h^\xi_\nabla r^2_s (\nabla x_{\text{pair}})^2] \,,\label{eq:Epair}
\end{eqnarray}
where 
\begin{equation}
    x_t=\frac{\rho_t}{\rho_{\text{sat}}}\,, \:  x_{\text{pair}}=\frac{\rho_0}{\rho_{\text{sat}}}\,,
\end{equation}
are dimensionless quantities, in which the pairing and saturation densities $\rho_\text{sat}=\rho_{\text{pair}}=0.16\text{ fm}^{-3}$ are fixed scaling parameters of Fayans EDFs. In these equations, we designate particle density $\rho_t$, spin-current density $J_t$, and pairing density $\tilde{\rho}_q$. In these energy densities, $q = n, p$ designate the proton and neutron particle types and $t = 0, 1$ specify the isoscalar and isovector densities. Furthermore, by comparison with the Skyrme EDF terms, the $\kappa$ and $g$ parameters can then be identified, which yields to $C^{\rho\nabla J}_0=K^{\rm ls}\kappa$, $C^{\rho\nabla J}_1=K^{\rm ls}\kappa'$, $C^{JJ}_0=K^{\rm ls} g$, and $C^{JJ}_1=K^{\rm ls} g'$, with $K^{\rm ls}=\frac{4\epsilon_F r_s^2}{3\rho_{\text{sat}}}$~\cite{Reinhard2017}.

It can be noted that, while this is not the case in our study, out of the six free parameters of the volume part of the Fayans EDF, namely $a^\pm_v,\ h^v_{1,2\pm}$, five can be directly expressed in terms of the following  nuclear matter properties: the equilibrium density $\rho_{\text{eq}}$, the symmetry energy $J$; the symmetry energy slope $L$, the energy per nucleon $E_B/A$ and the nuclear incompressibility $K$. This recoupling would only leaves $h^v_{2-}$ as a direct volume parameter, while replacing other model parameters by more commonly used physical constants in the droplet model~\cite{Reinhard2024}.

One of the most crucial points of Fayans functionals lies in their gradient of densities, namely in the normal and pairing parts of the Fayans EDF, where gradient of the normal density is present. For example, the $(\nabla x_{\text{pair}})^2$ term in Eq. (\ref{eq:Epair}) in essence simulates the effects of longer-ranged interactions, and non-local effects, and has been shown in earlier studies to present a stronger influence on the odd-even effects on the nuclear charge radius~\cite{Reinhard2017}. When computing the mean field, such term  outputs a term containing Laplacian of matter density, which peaks on the nuclear surface. As such, while the pairing correlations will be stronger in even-even nuclei, they will grow weaker on odd-$A$ nuclei due to the impact of quasiparticle blocking, impacting then via this term on mean-field on nuclear surface. This potential improvement was one of the main points regarding Fayans EDFs as this effect on the nuclear radii along the isotopic chains has been underestimated by most commonly used EDFs~\cite{Shi2020}.

Using the original Fayans EDF FaNDF0 parametrisation as a base, new parameter adjustment was carried out in~\cite{Reinhard2017}, utilizing data on various semi-magic and double-magic nuclei. In order to optimize the Fayans parameters, binding energies, charge radii, single-particle energy differences, neutron odd-even and even-even staggering energy differences were used. This lead to a new version of the reoptimized Fayans parameters, Fy(std), that would also be the foundation of subsequent Fayans EDFs. Down the line, those parameters were then readjusted using Calcium isotopes differential charge radii as additional constraints, which lead to Fy($\Delta r$). As initially observed in Ref.~\cite{Reinhard2017}, the parameter $h^\xi_\nabla$ is the key element driving the odd-even charge radius effects in Fayans calculations due to its implication on density gradients, as seen in Eq. (\ref{eq:Epair}). Thus, due to Fy(std) functional's optimization data constraining poorly the $h^\xi_\nabla$ parameter, its value became rather small, and, as a result, Fy(std) does not display any significant odd-even effect on the predicted Calcium isotopes charge radii. On the other hand, with Fy($\Delta r$), additional data on differential mean-square charge radii in Calcium isotopes was used, resulting to larger value of $h^\xi_\nabla$, and to larger odd-even effect in the charge radii.
Later on, a new adjustment of Fy($\Delta r$) within the HFB framework provided the Fy($\Delta r$,HFB) parametrization~\cite{Miller2019}, which allowed the new functional to operate on weakly bound nuclei. Recently, those Fayans functionals have been successfully applied on various other isotopic chains in connection with state-of-the-art charge radii measurements~\cite{Reinhard2020,Koszorus2021,Reponen2021,Kortelainen2022,Tolokonnikov2015}.

\section{Methodology}
The main objective in this work was to study the prevalence of octupole deformation within the Fayans EDF framework. Since the earlier studies with Skyrme EDFs have mapped a presence of octupole deformed nuclei in the actinide region, we decided to span our calculation to the region of $Z=82$ to $Z=100$ and $N=126$ to $N=142$. For the DFT calculations, we have utilized HFBTHO program~\cite{Stoitsov2013}, which solves the HFB equations by using axially symmetric harmonic oscillator basis. As such, its version 2.00d-fayans, affiliated to the 2.00 main release, and including the Fayans EDF implementation, was chosen for the calculations. As the used Fayans functionals were originally adjusted in coordinate space based approach, their pairing strength parameters needed to be readjusted due to different quasiparticle level density in the continuum when using oscillator basis based approach. In this work, we use the same pairing strength parameters for Fy(std) and Fy($\Delta r$,HFB) as in~\cite{Reponen2021}, which were adjusted to empirical pairing gap data of mid-shell Palladium isotopes.

In order to avoid missing any eventual significant octupole-deformed nuclei, we did systematic constrained calculations, allowing us to determine if the specified nucleus presents a non-zero octupole deformation through every nucleus in the proton number and neutron number directions centered around the octupole deformed nuclei cluster on the nuclear chart until nuclei displaying no octupole deformation was reached. Processing through the cluster with this method, we found the lower and upper neutron and proton number limits of the octupole-deformed actinides cluster. Focusing solely on the actinide region of the nuclear chart, we ultimately obtained a cluster contained within a square spanning $Z=84$ to $Z=108$ and $N=120$ to $N=150$, encapsulating all the nuclei of our theoretical cluster presenting a nonzero octupole deformation.

On a side note, a more comprehensive set of dimensionless parameters regarding the multipole deformation are deformation parameters $\beta_n$, which can be computed from the corresponding multipole moments. If those moments are expressed in powers of barns, we can, for example, define the quadrupole deformation parameter as
\begin{equation}
    \beta_2=(Q_2\sqrt{5\pi})/(3A^\frac{5}{3}R_0^2) \,,
\end{equation}
with $A$ being the mass number and $R_0$ the nuclear radius parameter expressed in $\text{b}^\frac{1}{2}$ ($R_0=0.12\ \text{b}^\frac{1}{2}$). Similarly, from the electric octupole moments expressed in $\text{b}^\frac{3}{2}$, we also defined the octupole deformation parameter as
\begin{equation}
    \beta_3=(Q_3\sqrt{5\pi})/(3A^2R_0^3) \,,
\end{equation}
or, more generally,
\begin{equation}
    \beta_n=(Q_n\sqrt{5\pi})/(3A^\frac{n+3}{2}R_0^n) \,.
\end{equation}

\subsection{Even-even nuclei}
The procedure for calculating even-even nuclei is following:
\begin{enumerate}
\item The upper and lower neutron and proton numbers bounds have been determined to encompass all octupole-deformed nuclei. For every nucleus, we performed a set of constrained calculations, with constrains on quadrupole $Q_2$ and octupole $Q_3$ moments. Additionally, center-of-mass was constrained to zero in all calculations.

\item For each even-even nucleus, we spanned the quadrupole moment from $Q_2=-25\text{ b}$ to $Q_2=35\text{ b}$ by steps of $1\text{ b}$ and the octupole moment from $Q_3=0\text{ b}^\frac{3}{2}$ to $Q_3=10\text{ b}^\frac{3}{2}$ by steps of $0.25\text{ b}^\frac{3}{2}$ through constrained calculations. The step sizes have been chosen so as to obtain sufficiently precise calculations across the $Q_2-Q_3$-plane. Whereas the former moment is an even-order deformation and so does not present symmetry between positive and negative values, the potential energy surface of an odd-order moment such as $Q_3$ is reflection-symmetric with respect of positive and negative values. In total, using both Fayans EDFs on each of the 16 isotopes of each of the 13 elements, we computed 1.040.416 independent self-consistent constrained calculations.

\item 
In order to compute the variational minimum, we performed unconstrained calculation while still keeping constraint on the center-of-mass. Upon convergence, we obtained the lowest point of the potential energy surface of the specified nucleus. Such process was applied to each of the 416 studied isotopes.
\end{enumerate}

\subsection{Odd-A nuclei}
It can be noted that the HFB state has a property called number parity. That is, HFB state describes a nucleus with an even number of protons and neutrons. However, by using the quasiparticle blocking procedure, one can also compute an odd-$A$ or odd-odd nuclei. This procedure will be further detailed below. The blocking procedure in itself is handled by using equal filling approximation~\cite{Martin2008}. The blocked state was selected from the set of low-lying quasi-particle states of the reference even-even nucleus. As such, we proceeded as follows:
\begin{enumerate}
\item In order to get these results for an odd-$A$ nucleus with $Z,N+1$, where $Z$ and $N$ are even, we first utilized the procedure described in previous section to compute neighbor even-even nucleus at $Z,N$. Then, from the set of available neutron quasiparticle state blocking candidates, we selected a few of the lowest-lying ones. 

\item Afterward, for each blocked Nilsson orbital, we performed an unconstrained calculation with the same initial starting values for $Q_2$ and $Q_3$ as before.

\item Finally, for each nucleus, we selected the blocked state which gave the lowest HFB energy, producing complete isotopic chains.

\end{enumerate}

Two additional aspects of quasiparticle blockings should be noted. Firstly, that the quasiparticle energy of the orbitals depends strongly on the initial and final deformation parameters. As such, it appeared to us inconsistent to operate the same extensive constrained calculations as for the even-even nuclei since the set of blocked quasiparticle states would most likely be based on a very different set Nilsson orbital. Thus, we have considered only the unconstrained calculations for the odd-$A$ nuclei, and thus have only compiled their ground state properties. Moreover, it can also be noted that odd-odd nuclei can be computed in the same fashion with the exception that we would now have to choose an orbital for both the proton and the neutron. This would then imply a much lengthier list of blocked quasiparticle state combinations to compute. In this work, we have limited our calculation to even-$Z$ isotopic chains, yet systematic calculations of odd-$Z$ nuclei would be an interesting topic for future studies.

\section{Results}
\subsection{Properties of even-even actinides}
\begin{figure}[htb]
\centering
\includegraphics[width=0.5\textwidth]{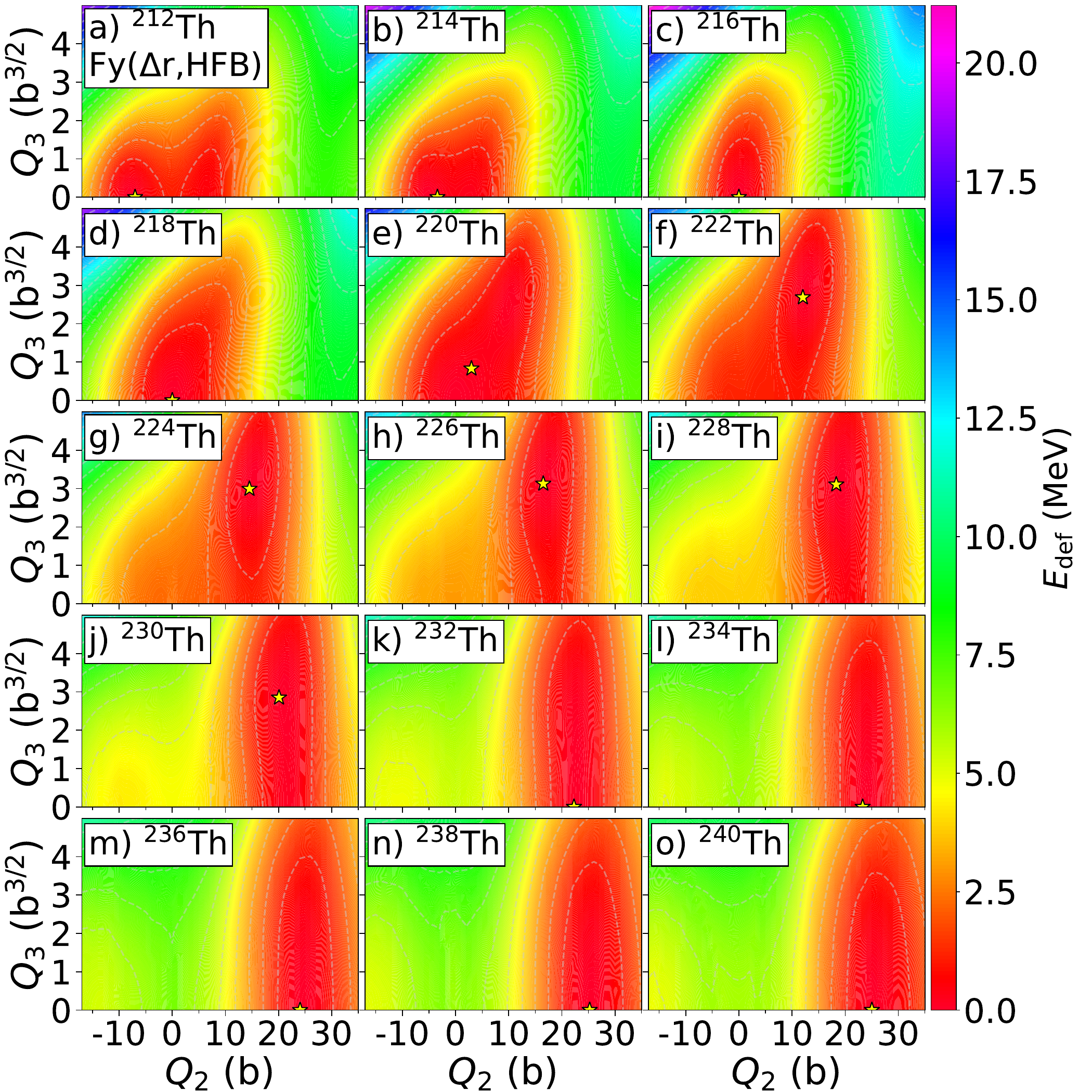}
\caption{Calculated ground state deformation energy of $122\leq N \leq150$ even-$N$ Thorium isotopes with Fy($\Delta r$,HFB), as a function of quadrupole and octupole deformations. The stars represent the potential energy minimum of each isotope obtained via the unconstrained calculations.}\label{fig:def1}
\end{figure}

On the following graphs on Fig. \ref{fig:def1}, we present the successive plots of deformation energy for one of the most demonstrative isotopic chain, namely the Thorium, for Fy($\Delta r$,HFB).

We can notice on Fig. \ref{fig:def1} a smooth transition between the ground state deformation energy surface of the neighboring nuclei across the isotopic chains. Moreover, at the beginning of the isotopic chain, the isotopes clearly do not present any octupole deformation, yet we can already observe from these potential energy surfaces that Fayans EDFs present a certain softness towards octupole deformation as we can observe a continuous valley on deformation energy across the vertical axis. This systematic valley of stability around the minimum ground state energy, just above 1 MeV deep at the maximum octupole deformation of the isotopic chain, was a crucial point that we expected to observe, as the Fayans functionals also aim to reproduce the results of Skyrme EDFs which have been documented to display such landscape~\cite{Robledo2013,Ebata2017}. Later on, the octupole deformation begins to show non-zero values for the ground state minimum for both functionals. This implies that the inclusion of octupole deformation in the calculations further lowers the ground state energy by an additional energy varying from 0.2 MeV for the lesser-impacted isotopes to up to 1.25 MeV for the most sensitive ones, for example, the isotopes at the middle of the chains.

Furthermore, as it can be observed at around $N=136$, the octupole deformation reaches its maximum for  Fy($\Delta r$,HFB) functional, hinting at the fact that non-zero octupole moment is an essential element with these nuclei. From this point onward, the octupole moment decreases, as it can be seen on the $N=140$ plot. Yet, despite being smaller than for its previous even-$N$ direct neighbor, it clearly appears that octupole deformation for those subsequent isotopes remains a non-negligible aspect. Finally, we reach the end of the non-zero octupole deformation at $N=142$. The Fy(std) functional exhibits similar kind of behavior. From this point on, neither of the functionals will predict a non-zero octupole moment. This point, where the isotopes revert back to a non-octupole-deformed state, varies from nuclear species to species. For example, the Thorium isotopic chain displays a non-zero octupole deformation up to $N=142$, while the Plutonium one continues until $N=148$.

\begin{figure}[H]
\includegraphics[width=\columnwidth]{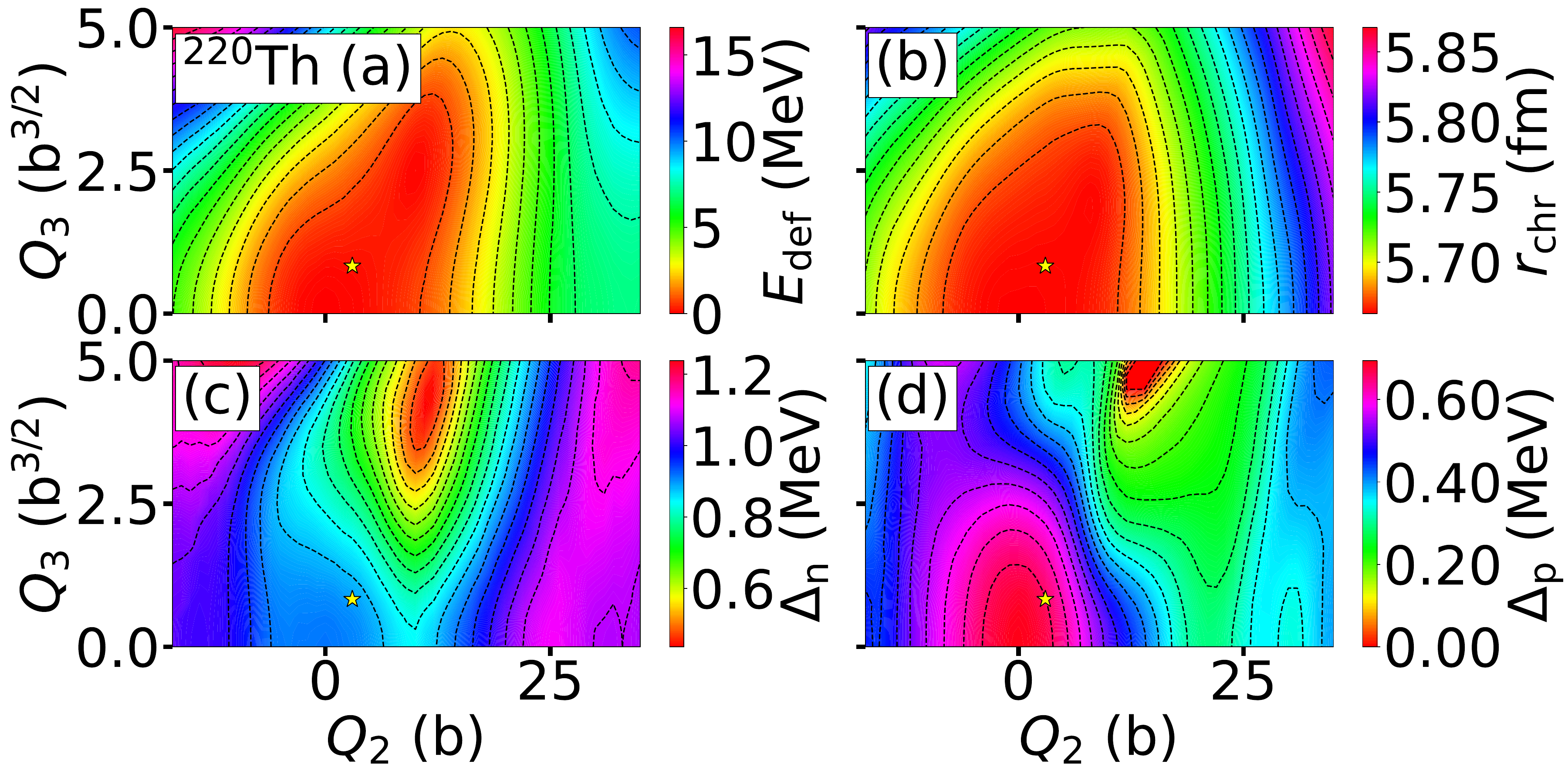}	\caption{Deformation energy in MeV (a), rms charge radius in fm (b) and both the neutron (c) and proton (d) pairing gaps in MeV in $^{220}$Th, as a function of quadrupole and octupole moment, computed with Fy($\Delta r$,HFB). Star symbol represents the potential energy minimum obtained via an unconstrained calculation.} \label{fig:four-props}
\end{figure}

Additionally, we also noted the notable impact of the octupole deformation in Fayans EDFs calculations on various other nuclear properties such as the nuclear charge radius or the pairing gaps. As it can be seen in Fig. \ref{fig:four-props}, where we have taken $^{220}$Th as an example, we can observe a distinct quadratic relation between the charge radius and the two multipole moments, as expected, as well as variations in the pairing gaps, indicating a strong responsiveness of the Fayans functionals regarding the local effects across the deformation landscape induced by octupole deformation.

Overall, this behavior highlights the expected significant impact a non-zero octupole parameter has on the ground state deformation energy, and thus puts in perspective the relevance of octupole deformation for actinides and trans-actinides isotopes. Furthermore, these results are in accordance with previous theoretical studies conducted with various different EDF models~\cite{Agbemava2016,Agbemava2017,Nomura2021a,Nomura2021b,Cao2020,Xia2017}. This tends to indicate that Fayans EDFs can be expected have a similar predictive power in octupole-deformed nuclei. This transition between non-octupole deformed and octupole deformed nuclei proceeds similarly for all calculated isotopic chains, with both Fayans EDFs, including the aforementioned effects caused by octupole deformation.

\begin{figure}[htb]
\includegraphics[width=\columnwidth]{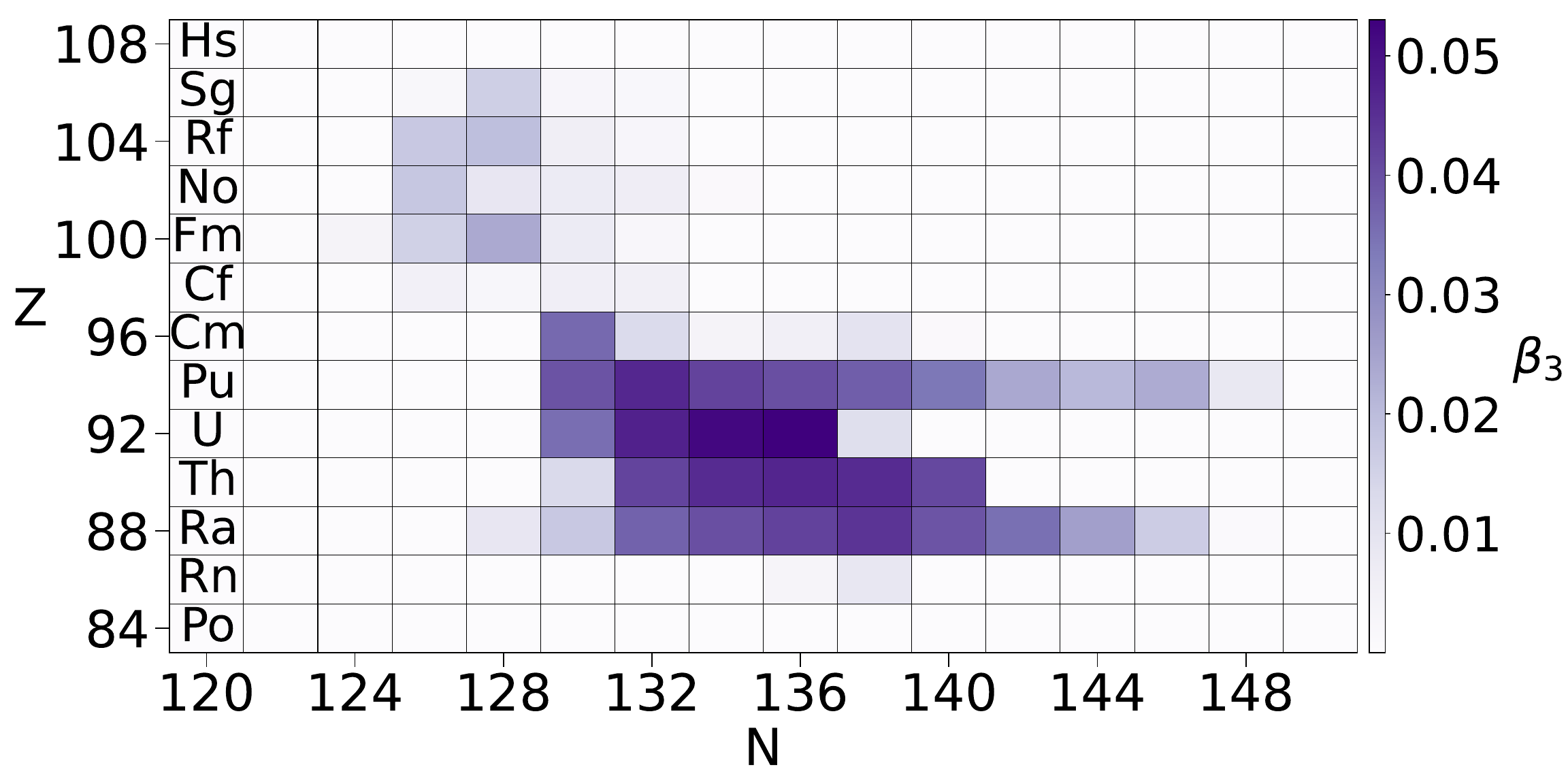}	
\caption{Octupole deformation parameter of even-Z isotopes within the whole actinides cluster computed with Fy($\Delta r$,HFB).} \label{fig:chart1}
\end{figure}

\begin{figure}[htb]
\includegraphics[width=\columnwidth]{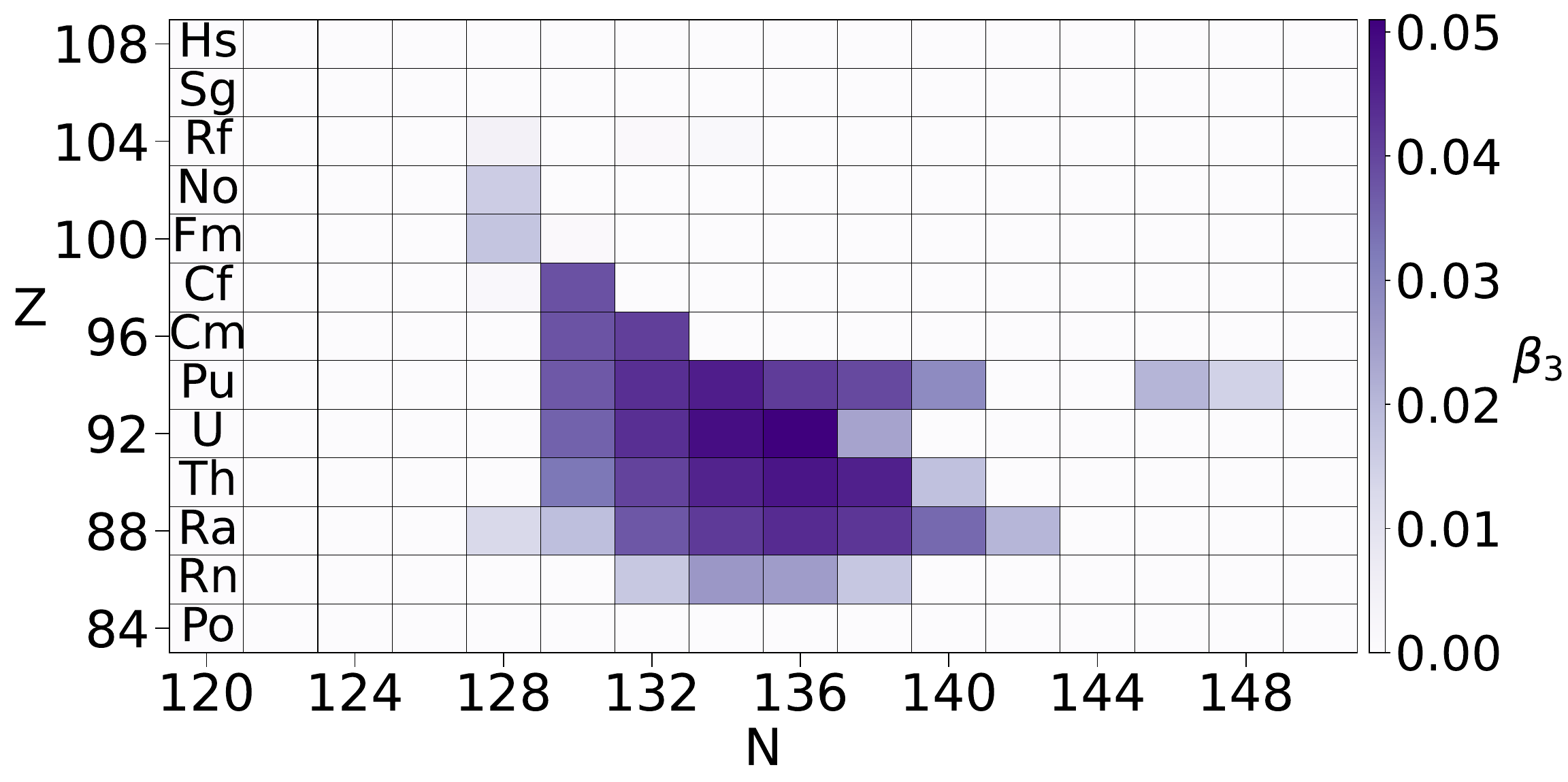}	
\caption{Same chart as in Fig. \ref{fig:chart1} but with Fy(std).} \label{fig:chart2} 
\end{figure}

On Figs. \ref{fig:chart1} and \ref{fig:chart2}, we show the $\beta_{3}$ value charts for the whole cluster for both functionals. As a reference for further comparisons with quadrupole moment calculations, the similar charts of $\beta_{2}$ deformation factors are available in our online repository~\cite{repo}.

\begin{figure}[htb]
\includegraphics[width=\columnwidth]{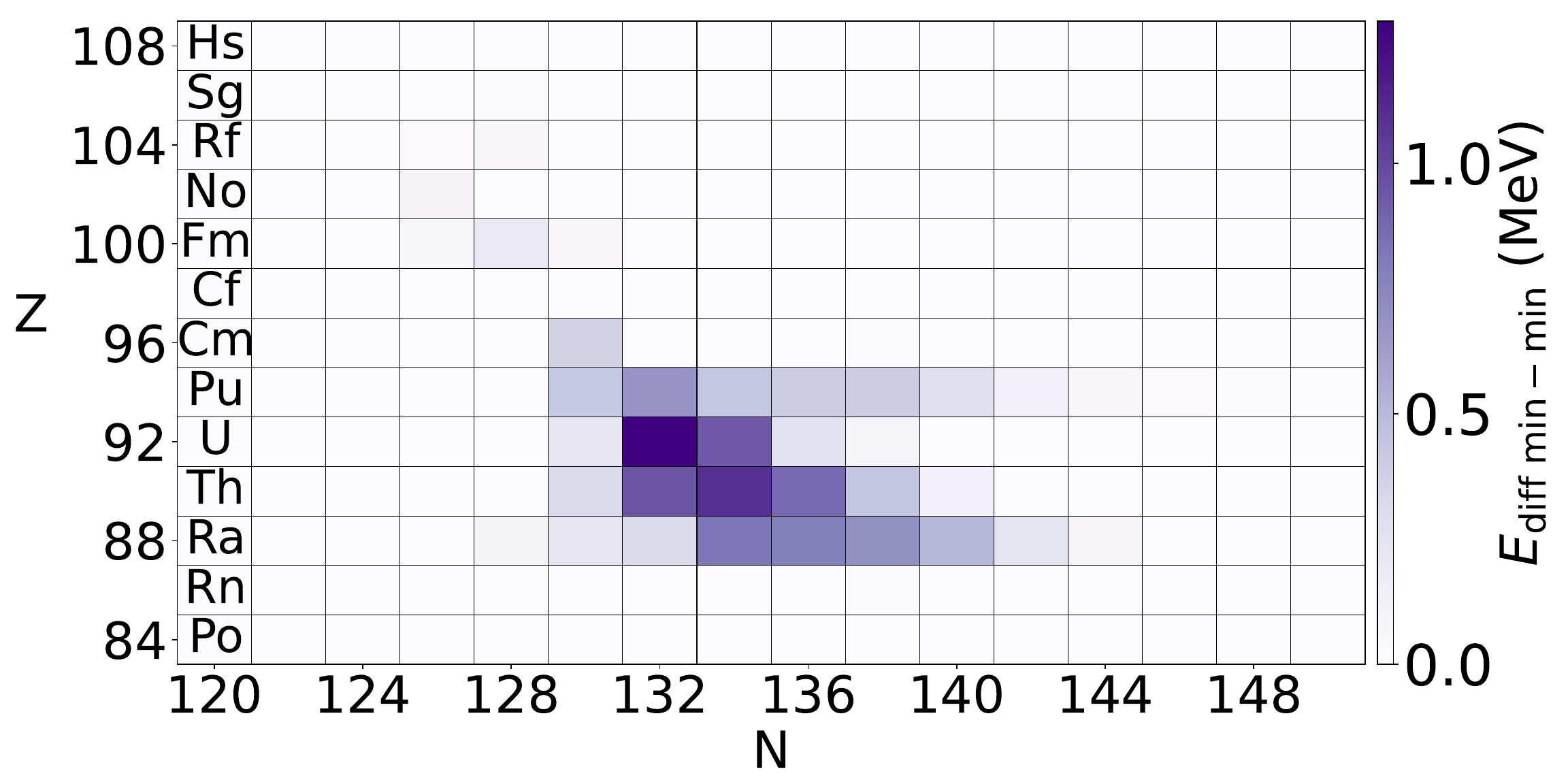}	
\caption{Ground state energy gained from the inclusion of a possible non-zero octupole deformation computed with Fy($\Delta r$,HFB).} \label{fig:chart5}
\end{figure}

\begin{figure}[htb]
\includegraphics[width=\columnwidth]{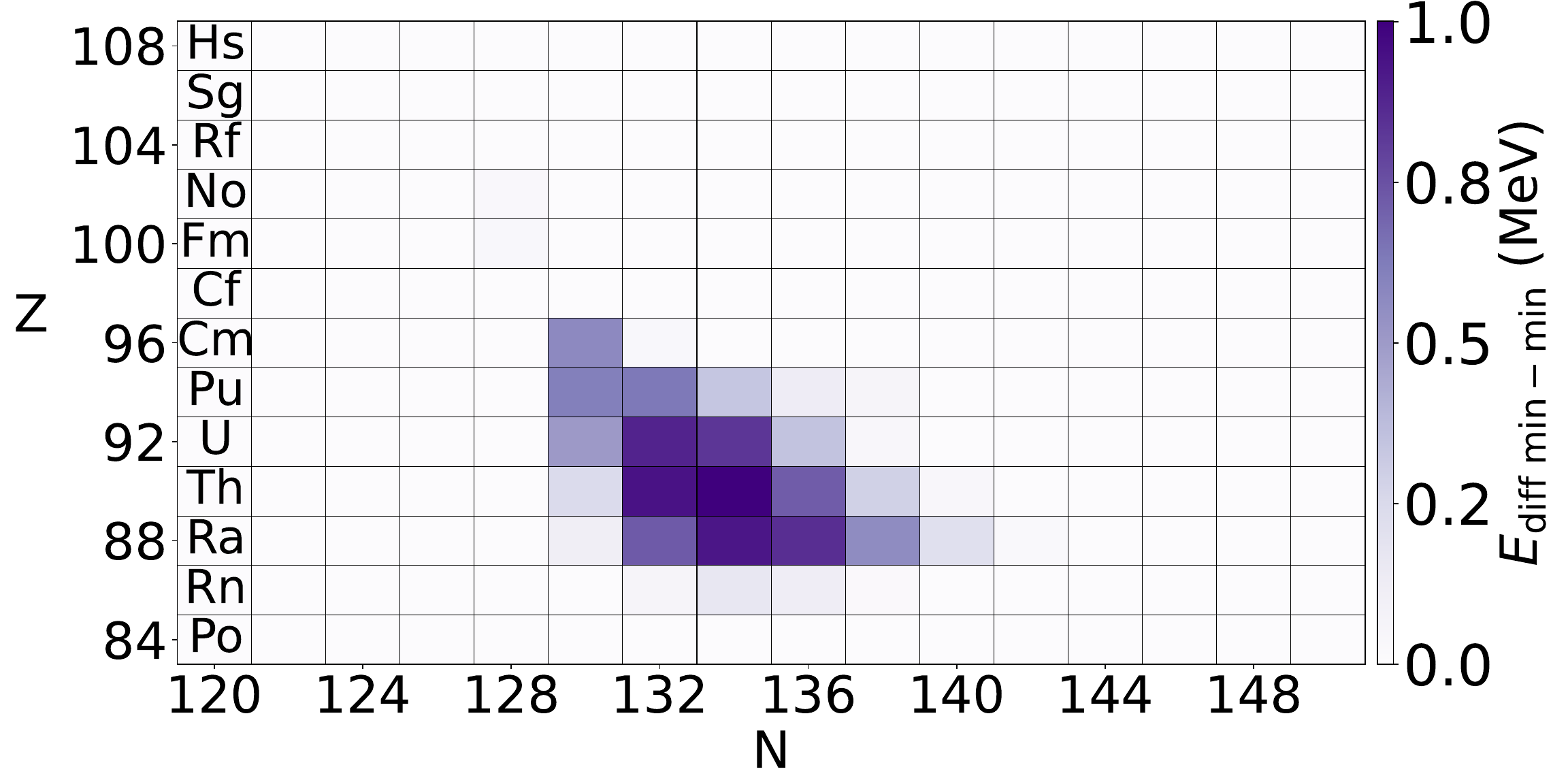}	
\caption{Same chart as in Fig. \ref{fig:chart5} but with Fy(std).} \label{fig:chart6}
\end{figure}

In addition, Figs. \ref{fig:chart5} and \ref{fig:chart6} show the ground state energy difference between the usual unconstrained computation with just the quadrupole factor and the unconstrained computation with both the quadrupole and octupole moments enabled.

The charts displayed on Figs. \ref{fig:chart1}--\ref{fig:chart2} clearly show a well-defined octupole cluster within the actinides' region, with a strong core and diluted edges. Furthermore, by comparing our Fayans results with the previous survey on pear-shape landscape using Skyrme-based EDFs~\cite{Cao2020}, the cluster brought out in this work appears to coincide both in location within the nuclear chart, as the cluster appears to be centered around $^{226}$U, in shape, as it stretches toward the neutron-heavy $84\leq Z\leq94$ nuclei and towards the proton-heavy $126\leq N\leq128$, and in the magnitude of the octupole deformation parameter value.

In addition to octupole deformation parameter, a similar region shape can be observed on Figs. \ref{fig:chart5}--\ref{fig:chart6}. These charts highlight the fact that octupole deformation is a highly relevant factor for these nuclei. As was indicated in earlier deformation energy plots with both functionals, the inclusion of octupole deformation generally can lower the binding energy by around 1 MeV. Additionally, the Fy($\Delta r$,HFB) EDF appears to predict slightly stronger octupole deformation as whole, with additional 0.2 MeV lower deformation energy as compared to Fy(std). It can be noted that, in both cases, only the center of the cluster, the area presenting a significant octupole deformation parameter, which threshold we have estimated at $\beta_3>0.02$, is significantly impacted by the octupole deformation in itself.

Both observations underline the fact these Fayans functionals predict the expected behavior as they appear capable of reproducing similar features as Skyrme-based EDFs, handling both quadrupole and octupole deformations with the expected relevancy. Finally, these results show that, in actinide region, the octupole deformation appears highly relevant with Fayans functionals, as it contributes to the overall nuclear shape stability by lowering the nuclear ground state energy minimum with a non-negligible amount.

\subsection{Isotopic chains of even-$Z$ nuclides and comparison to UNEDF0 EDF}
\begin{figure}[htb]
\includegraphics[width=0.95\columnwidth]{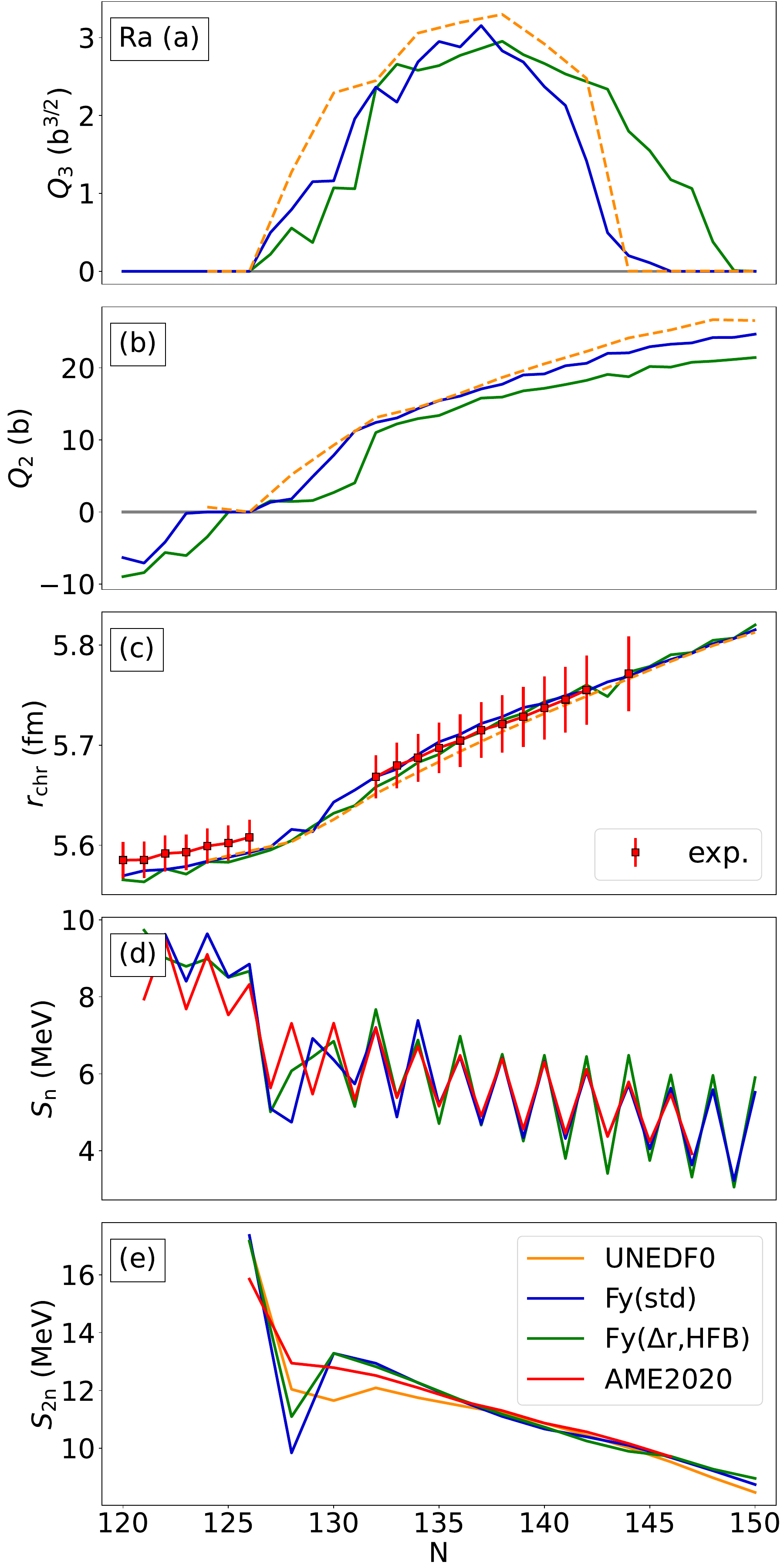}
\caption{Octupole moment (a), quadrupole moment (b), the rms charge radius (c), the neutron separation energy (d), and the two-neutron separation energy (e) across the Radium isotopic chain with Fy($\Delta r$,HFB) and Fy(std), compared to even-$N$ UNEDF0 EDF predictions (indicated with dashed line in panels (a)-(c)), and experimental data on radii and separation energies when relevant.}\label{fig:chain1}
\end{figure}

\begin{figure}[htb]
\includegraphics[width=0.95\columnwidth]{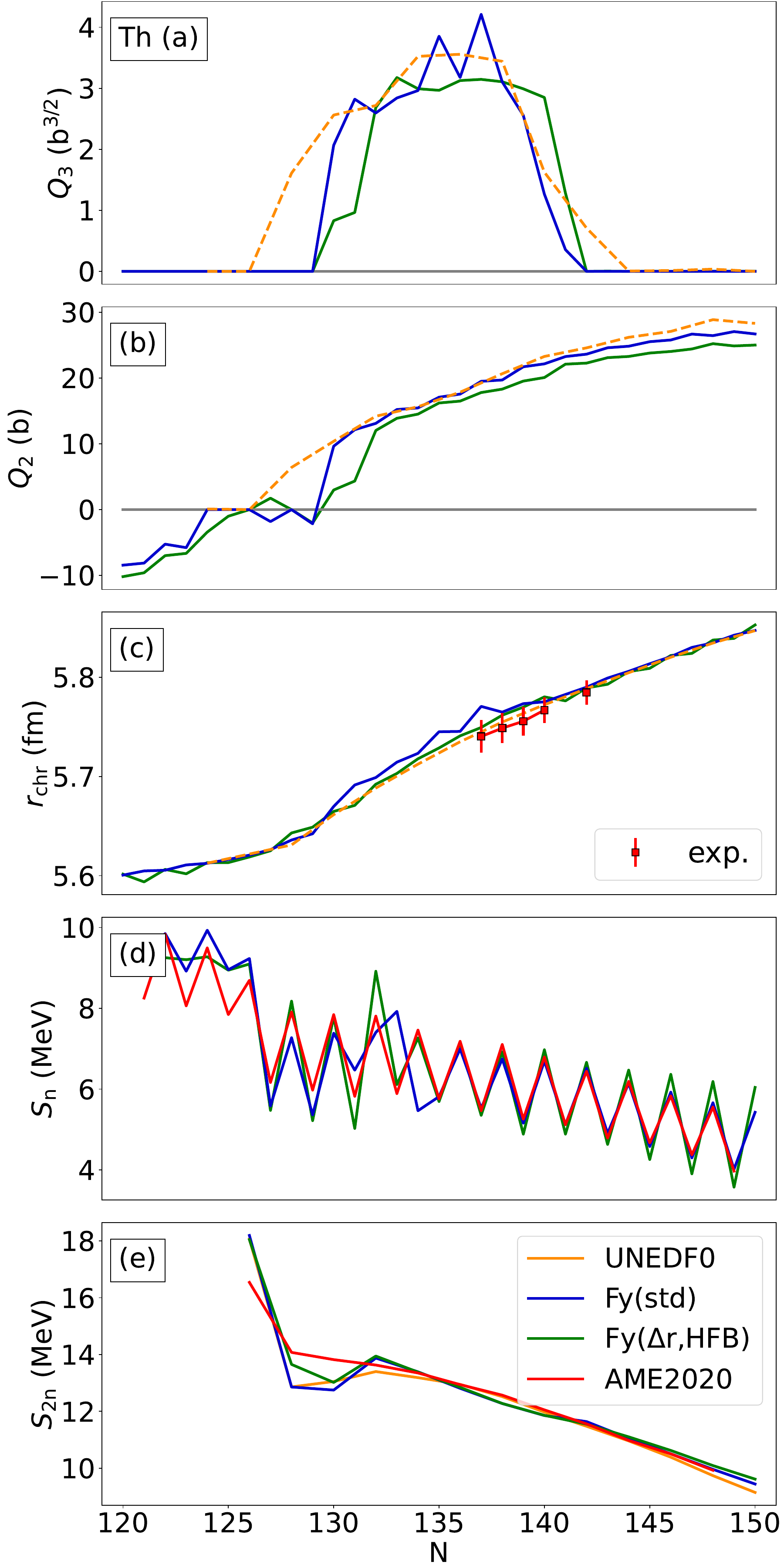}
\caption{Same plots as in Fig. \ref{fig:chain1} but for the Thorium isotopic chain.}\label{fig:chain2}
\end{figure}

\begin{figure}[htb]
\includegraphics[width=0.95\columnwidth]{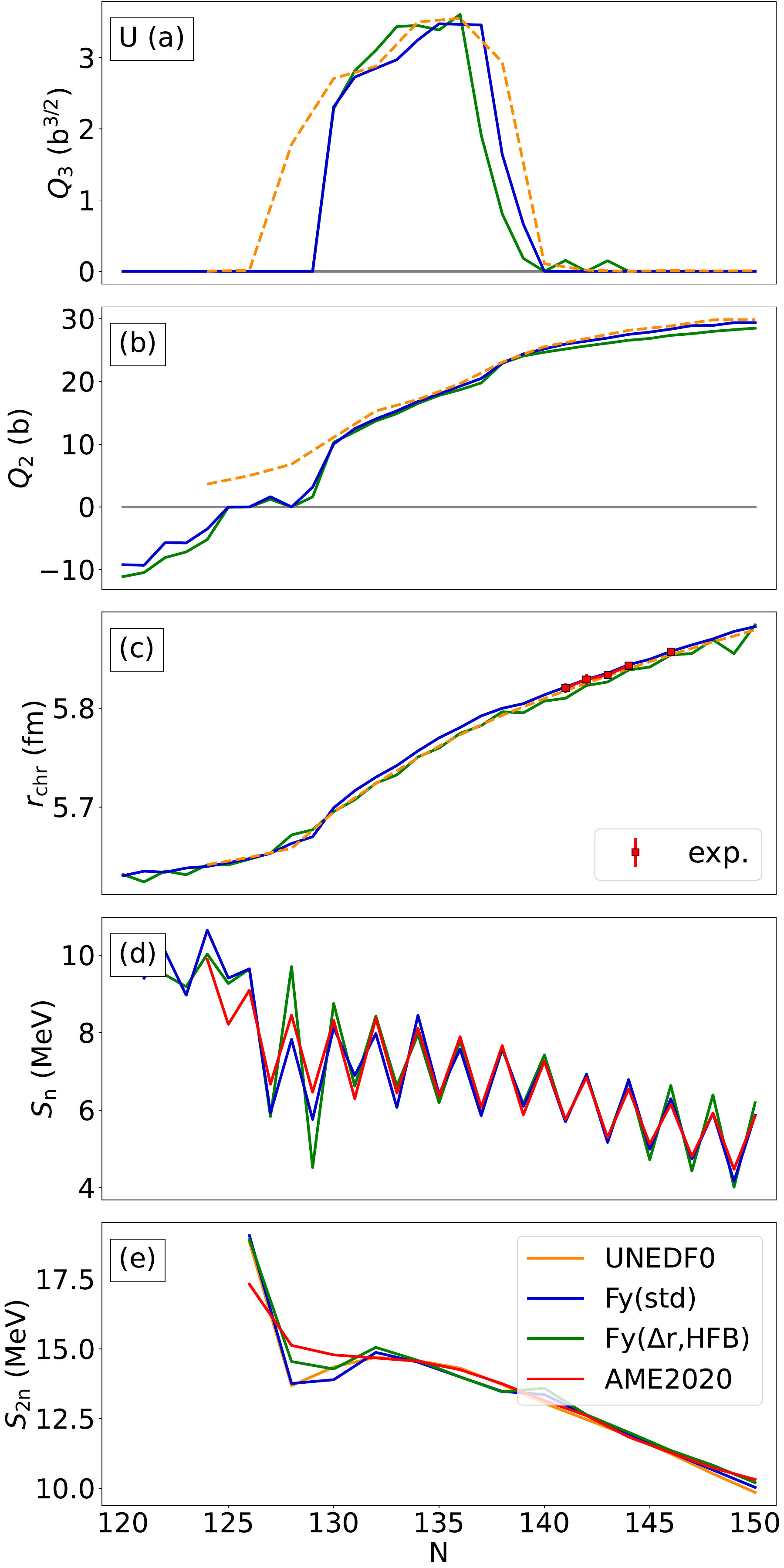}
\caption{Same plots as in Fig. \ref{fig:chain1} but for the Uranium isotopic chain.}\label{fig:chain3}
\end{figure}

\begin{figure}[htb]
\includegraphics[width=0.95\columnwidth]{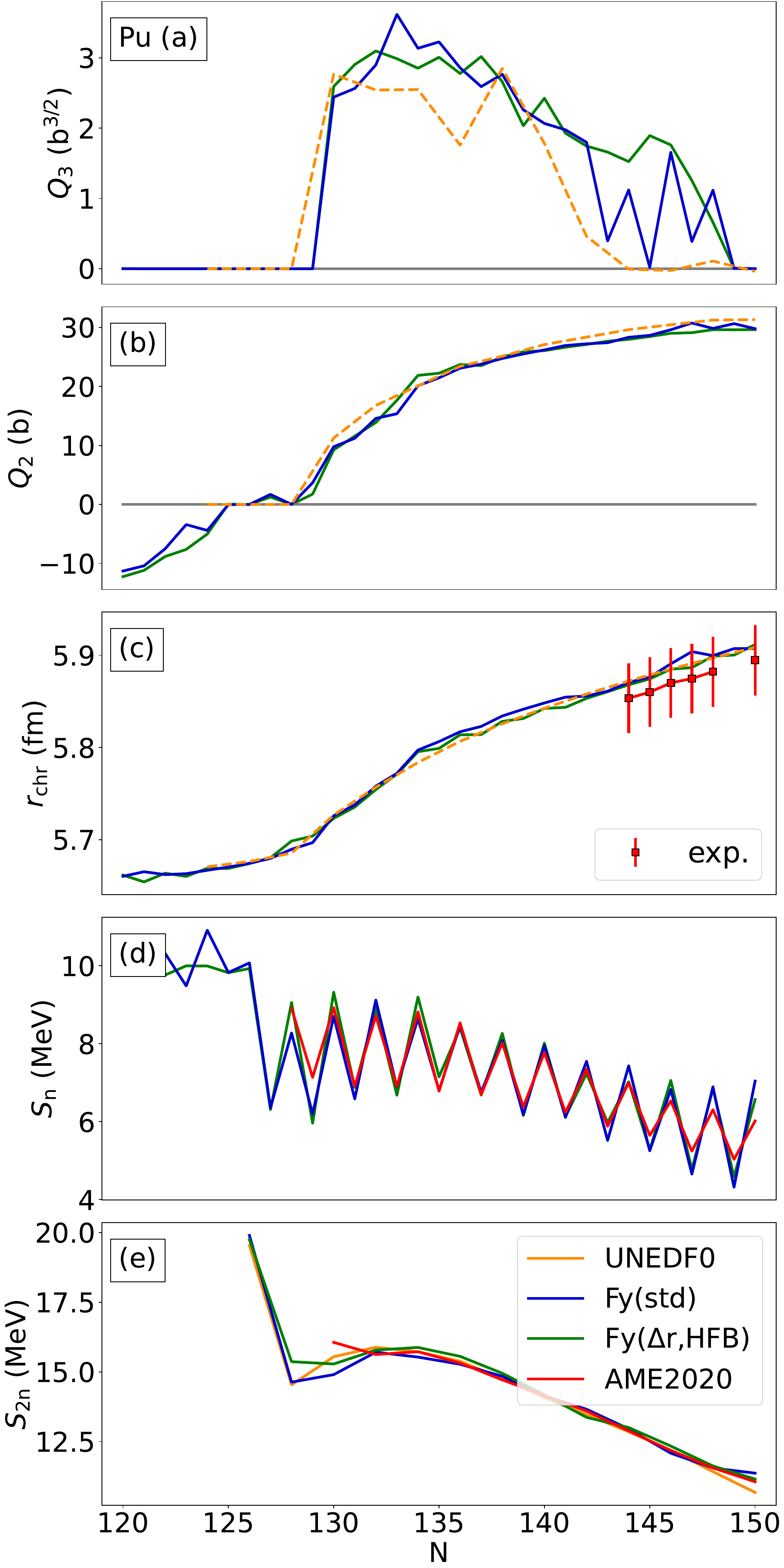}
\caption{Same plots as in Fig. \ref{fig:chain1} but for the Plutonium isotopic chain.}\label{fig:chain4}
\end{figure}

As the next step in this work, we have studied the behavior of quadrupole and octupole moments, rms charge radius, and one and two-neutron separation energies throughout the whole isotopic chains. These nuclear properties are shown in Figs. \ref{fig:chain1}--\ref{fig:chain4}, displaying Radium, Thorium, Uranium and Plutonium chains for both Fayans functionals. Alongside, are plotted the even-even UNDEF0 calculations as a benchmark, which will be discussed later on. Furthermore, as a point of comparison, the experimental nuclear radii extracted from a combined experimental survey employing both isotopic shift-induced radii changes and electronic scattering measurements for the absolute radii~\cite{Angeli2013} and high-resolution laser spectroscopy measurements of heavy nuclei spectra~\cite{Campbell2016, Lynch2018, Verstraelen2019}, as well as the experimental neutron separation energies extracted from the Atomic Mass Evaluation 2020 database~\citep{ame20}, have been plotted alongside the computed ones. In panels (a), (b), and (c) of Figs. \ref{fig:chain1}--\ref{fig:chain4}, the dashed UNEDF0 lines specify the fact that, in these panels, they include even-$N$ nuclei only, contrarily to the other solid lines in those same plots, which include also odd-$N$ nuclei.

Overall, both Fayans EDFs appear to produce the expected behavior regarding the four selected nuclear properties, with no unnatural kink in the graphs, indicating a proper simultaneous handling of octupole deformation and quasiparticle blocking in the calculations.

Regarding the octupole and quadrupole deformation, both functionals appear to display a rather smooth bell-shaped curve along the isotopic chains, and an expected behavior for the quadrupole deformation when crossing the $N=126$ shell gap. Across each of the $84\leq Z\leq94$ isotopic chains presented in Figs. \ref{fig:chain1}--\ref{fig:chain4}, one can observe how, starting from a non octupole-deformed nucleus, $\beta_3$ deformation manifests itself, reaches a peak of octupole deformation before diminishing and finally setting back to a nuclei displaying only reflection symmetric shapes, indicating a distinct island of octupole deformation within the actinides cluster. Furthermore, with the charge radii, the Fy(std) EDF appears to show a relative smooth increase along the isotopic chain, whereas Fy($\Delta r$,HFB) tends to display a more notable odd-even effect. As discussed earlier, this is linked to the construction of the Fy($\Delta r$) and Fy($\Delta r$,HFB) EDFs~\cite{Reinhard2017}.  Finally, both Fayans EDFs seemingly reproduce the general trend in neutron separation energy well, with some minor exceptions.

Overall, both Fayans EDFs appear to reproduce the experimentally observed isotopic trends of charge radius. However, as it can be observed on the Thorium chain on Fig. \ref{fig:chain2}(c), while they present a visually striking odd-even effect, which is an element that Fayans EDFs sought to achieve better than their Skyrme counterparts, the Fayans EDFs present opposite tendencies. Moreover, this odd-even effect in itself appears unequally across the isotopic chains, and in notably different aspects for each EDF. With Fy($\Delta r$,HFB), we have obtained larger radius for even-$N$ nuclei almost exclusively at the beginning and end of the chain. Meanwhile, with Fy(std), odd-$N$ nuclei display a larger radius, and having odd-even effect comprised only at the central part of the isotopic chain. However, given its parameters, it was a surprise to observe such large odd-even effects with Fy(std) in actinides isotopes, as the small magnitude of pairing parameter $h^\xi_\nabla$ was observed to produce almost no odd-even effects on medium-weight isotopic chains~\cite{Reinhard2017}. Nevertheless, the presence of this staggering clearly highlights the stronger aforementioned effects in heavy actinides originating from the gradient term in pairing part of the Fayans EDF as compared to Skyrme-based EDFs.

Our calculated charge radius staggering appears often inverted as compared to normal case where, on average, the rms radius of an odd-$N$ nucleus tends to be smaller than the average of its even-$N$ neighbors. Such behavior, inconsistent at first glance, could actually be crucial element in this survey as previous experimental surveys using resonance ionization spectroscopy hinted at such an inverted staggering, linked with a reflection-asymmetric shapes in actinides~\cite{Verstraelen2019}. Furthermore, comparison to experimental results tend to indicate a varying degree in accuracy of the models. As it can be observed on the Thorium chain, at the beginning of the Radium chain, or in the Radon chain as displayed in our online database, some calculated charge radii lie significantly far from the experimental measurements range, including the experimental error bars. However, as it can be observed in the central part of the Radium chain, or in the Plutonium chain, both Fayans EDFs predict charge radii within the experimental error bars. Nevertheless, it must also be noted that the particularly accurate Uranium charge radii measurements are nearly perfectly reproduced, both in values and staggering, by the Fy(std) EDF. However, it can be overall observed that, apart from a few cases such as the Uranium chain, Fy($\Delta r$,HFB) predicts the adequate odd-even staggering while Fy(std) predicts an opposite trend for the staggering. Overall, based on the results predicted by both Fayans EDFs, it can be said that these EDFs still require extensive improvements, as some radii are not reproduced accurately enough, although experimental data on some isotopic chains is excellently reproduced.

Lastly, we can clearly observe how both Fayans EDFs reproduce accurately both the general trend and the odd-even effect in neutron separation energy when compared to the experimental measurements. Such an example can be seen on Uranium chain on Fig. \ref{fig:chain3}(d), where the experimental values and tendency are almost perfectly reproduced by both Fayans EDFs on $135\leq N\leq144$ nuclei. This indicates that, in addition to being consistent with each other, the nucleon pairing correlation of these EDFs are phenomenologically satisfactory and adequate in their respective adjustments. However, it can also be noted that Fy($\Delta r$,HFB) appears to deviate from the experimental values more strongly and frequently than Fy(std), while Fy(std) was seen to miss some neutron pairing gaps more often than Fy($\Delta r$,HFB), as we can observe on Fig. \ref{fig:chain1}(d) for $126\leq N\leq130$ and Fig. \ref{fig:chain2}(d) for $132\leq N\leq134$. This source for this inconsistency may be linked either deficiency in the pairing interaction or the lack of the time-odd channel in the EDF itself. Finally, these results corroborate the Fayans functional as a valid alternative to Skyrme-based EDFs.

Next, we compare predictions of the Fayans EDFs to the ones produced by the UNEDF0 Skyrme EDF. As the adjustment process of UNEDF0 included a large dataset of several deformed nuclei, it can be considered as a well-suited benchmark EDF with a good predictive power on heavy deformed nuclei. Using a similar methodology, we carried out unconstrained calculations on the isotopic chains with both quadrupole and octupole deformations enabled. We carried out UNEDF0 calculation for even-even nuclei only. This implies that only the two-neutron separation energy will be compared between Skyrme and the previously obtained Fayans-based predictions, while excluding one-neutron separation energy.

From Figs. \ref{fig:chain1}--\ref{fig:chain4}, we can clearly see how the three quantities on (b), (c), and (e) panels, predicted with Fayans EDFs, all coincide within a reasonable margin between the experimental separation energies and charge radii, as well as with the predictions by the UNEDF0 EDF, with only mostly minor differences between Fy($\Delta r$,HFB) and the other two EDFs on lighter isotope's quadrupole deformation. This indicates that the Fayans EDFs can reproduce the behavior of the UNEDF0 EDF, which has been rather successfully benchmarked against various experimental data sets of nuclear masses and radii. Furthermore, it must be noted that the Fayans EDFs behavior regarding octupole deformation, as seen on panels (a), differs somewhat from the UNEDF0 one, which tends to display a stronger intrinsic octupole moment spanning a shorter array of isotopes. In addition, looking only at the even-$N$ points, the former results appear to present a lesser number of abrupt kinks along the isotopic chain, especially in Fy($\Delta r$,HFB). Nevertheless, further experimental data, which would help to deduce octupole deformability, are needed in order to properly assess the accuracy of these Fayans and UNEDF0 calculations regarding such matters.

Overall, it appears that the two Fayans functionals not only replicate the UNDEF0 results, but also predict a consistent octupole deformation between each other as well as devoid of major kink, especially Fy($\Delta r$,HFB). In the end, we can conclude that these two Fayans functionals reproduce with significant success both the theoretical results of the previous survey conducted with Skyrme-based EDFs and the experimental data regarding the quadrupole landscape. All the computed individual plots, four properties plots, charts, isotopic chains and comparison graphs are available in our repository~\cite{repo}.

\section{Conclusion}
Through this survey, it was shown that Fayans EDFs can provide accurate results over the selected actinide cluster when compared to the reference current-generation Skyrme EDFs regarding quadrupole deformation, two-neutron separation energy, and nuclear rms radii. As such, Fayans functionals can be expected to reproduce most of the crucial aspects of Skyrme-based EDFs, such as UNEDF0, with high accuracy while at the same time having improved description of the local variation of the charge radius. Two main accomplishments of Fayans EDFs, as shown in this survey, are their relatively accurate prediction of charge radius, including the odd-even effect, as well as their octupole deformation properties and its impact of the various analyzed nuclear observables and their systematics. As highlighted earlier, our results showed that Fayans EDFs have similar octupole deformation properties as the Skyrme-based EDFs in actinide region, predicting a rather similar cluster of octupole deformed nuclei as obtained in earlier studies. Therefore, it is clear that octupole deformation cannot be neglected when Fayans EDFs are applied to actinide nuclei for high-precision predictions. As such, Fayans functionals can offer relatively accurate separation energies and radii predictions in heavy deformed nuclei.

Comparison to the previous DFT studies on octupole deformation properties shows that these Fayans EDFs behave similarly, paving a way towards applying Fayans EDFs to complex related properties or processes, such as the Schiff moment or fission. Moreover, in addition to being consistent with earlier Skyrme EDF studies, these Fayans EDFs also predicted more minute new effects regarding local variation of the charge radius along the isotopic chain, or the stronger preponderance of octupole deformation. 
Additionally, we can note the Fy(std) results on Uranium isotopes, which turned out to be particularity accurate when compared to experimental results. 
As such, Fayans EDFs appear to successfully reproduce the aspects the Skyrme EDFs are known for, such as their accuracy with various bulk properties, while at the same time expanding beyond this scope for example with the nuclear charge radii.

In the end, this systematic survey of octupole deformation helped to demonstrate how both the Fy($\Delta r$,HFB) and the Fy(std) EDFs can be considered as adequate starting points in the work towards new generation of EDFs, intended for single-reference calculations, as a successor of the current Skyrme-based EDFs.
Furthermore, as the intrinsic octupole moment is considered to be an essential aspect in many nuclear properties, Fayans EDFs can also be considered as a great candidate for future similar surveys.

As a final note, this survey also put into perspective the accomplishments of these two Fayans EDFs, as it was shown that they also still present inaccuracies when compared to experimental data. 
As such, there still remains many possibilities for further research. Most applications of Skyrme EDF use different pairing strength for protons and neutrons. A recent development in Ref.~\cite{Reinhard2024} showed that such approach  generally improved the agreement of the Fayans EDF model with nuclear ground-state properties. Additionally, adjustment of Fayans EDF parameters at deformed HFB level could improve predictive power in deformed nuclei. In addition, the development process of new Fayans EDFs would also benefit from extensive surveys of current Fayans EDF properties across the whole nuclear chart, including odd and odd-odd nuclei.

\section*{Acknowledgments}
We would like to thank J. Dobaczewski for various helpful discussions. This work was possible thanks to the funds granted by the Research Council of Finland (Grant No. 339243, and the Centre of Excellence in Neutron-Star Physics, Grant No. 374066). We acknowledge the CSC-IT Center for Science Ltd. (Finland) for the allocation of computational resources.

\bibliography{fayan}
\end{document}